\documentclass{aastex}

\slugcomment{DRAFT - To appear in AJ}

\shortauthors{Ardila, Mart\'\i n, Basri}
\shorttitle{Low-mass Members of Upper-Scorpius}

\begin{document}

\newcommand \vsini{$v$sin$i$~}
\newcommand \caII{\ion{Ca}{2}~}
\newcommand \heI{\ion{He}{1}~}
\newcommand \liI{\ion{Li}{1}~}
\newcommand \hal{H$\alpha$~}
\newcommand \hbeta{H$\beta$~}
\newcommand \hgam{H$\gamma$~}
\newcommand \rsun{R_\odot}
\newcommand \msun{$\rm{M_\odot}$}
\newcommand \lsun{L_\odot}
\newcommand \kms{km s$^{-1}$~}
\newcommand \degs{$^\circ$}

\title{A Survey for Low-mass Stars and Brown Dwarfs in the Upper-Scorpius OB Association}

\author{David R. Ardila\altaffilmark{1}, Eduardo L. Mart\'\i n\altaffilmark{2}, Gibor Basri\altaffilmark{1}}
\altaffiltext{1}{Astronomy Dept., Univ. of California, Berkeley, CA 94720,\\e-mail: ardila@garavito.berkeley.edu, basri@soleil.berkeley.edu}
\altaffiltext{2}{Division of Geological and Planetary Sciences, Caltech 150-21, Pasadena, CA 91125,\\e-mail: ege@gps.caltech.edu}

\begin{abstract}
The Upper-Scorpius association is the OB association nearest to the Sun (145 pc). Its young age ($\sim$5 Myr) makes it an ideal place to search for low-mass stars and brown dwarfs, as these objects should be relatively bright. We have performed a photometric search for the low-mass members of the association, using the R, I, and Z filters. The completeness limit is $I\sim18.5$ and the saturation limit is $I\sim13$. We obtain 138 candidate members, covering nearly the entire M spectral type range. We find an excess of brown dwarf candidates over the number predicted by a Miller-Scalo Initial Mass Function. In addition, we have performed infrared imaging and low resolution optical spectroscopy of selected candidates. We find that the infrared observations confirm the spectral types obtained with the optical photometry. Furthermore, we find \hal in emission in 20 of the 22 objects observed spectroscopically. As \hal is an indicator of youth, we believe that these 20 objects may belong to the association. One of them, UScoCTIO 128 has a very strong and constant \hal line (equivalent width: -130 \AA), and its position in the color magnitude diagram suggests that it is a brown dwarf with mass equal to 0.02 \msun. Confirmation of this and the other candidates will have to wait for higher resolution observations that can reveal spectroscopic mass indicators like Li I and gravity indicators, such as K I and the subordinate lines of Na I.

\end{abstract}


\section{Introduction}
OB associations and gravitationally unbound clusters are likely to be the 
dominant birthplaces for the low-mass field star population (Preibisch \& Zinnecker 1999). Furthermore, they
 provide an opportunity to study very young low-mass stars and brown dwarfs, 
since these objects should be relatively bright in very young regions. Besides their intrinsic interest, the low-mass population of OB associations can provide constrains on the shape of the Initial Mass Function (IMF) at low masses. There are indications (Bouvier et al. 1998) that the shape of this IMF may deviate from the simple Miller-Scalo (Miller \& Scalo 1979) at masses less than 0.1 \msun.

While the low-mass stellar content of T associations (such as Taurus-Auriga, see Brice\~{n}o et al. 1998 and Kenyon 
\& Hartmann 1995) or very young clusters (such as IC 348, see Herbig 1998) seems 
well known, not much is known about the low-mass stellar content of OB 
associations. Part of the difficulty is due to the fact that most of the low-mass pre-main sequence 
(PMS) stars  cannot be easily distinguished from normal field stars in the 
huge area on the sky (several hundred square degrees) covered by nearby OB 
associations. Only Classical T-Tauri stars can be found easily by their 
strong \hal emission, using objective-prism surveys. X-ray 
observations have also proved to be an efficient way of distinguishing PMS stars from older field stars (see below).

The Scorpius-Centaurus association is the OB association nearest 
to the Sun. It contains several hundred B stars arranged in three 
subgroups: Upper-Scorpius, Upper Centaurus-Lupus and Lower 
Centaurus-Crux. Upper-Sco ($l\simeq354^o$, $b\simeq20^o$) is 
the youngest subgroup. Its lack of dense molecular material and 
deeply embedded young stellar objects indicates that the 
process of star formation has ended. The area is free of dense gas 
and clouds, and the association members show only moderate extinction 
($A_V<2$ mag).

Hipparcos data has been used to identify 120 association members, 
including 49 B stars and 34 A stars. According to these data, 
Upper-Sco is 145 pc away from the Sun and has a size of  $\sim130 
\rm{deg}^2$ (De Zeeuw et al. 1999).  Age determinations based on 
the upper (de Zeeuw \& Brand 1985; de Geus et al. 1989) and lower 
(Preibisch \& Zinnecker 1999) mass range derive a value of 5 Myr. However, there have been age determinations that seem to suggest that many of the low-mass cluster members could be 10 Myr old (Mart\'\i n 1998; Frink 1999). 

The first search for low-mass members of Upper-Sco was performed 
by Walter et al. (1994). They obtained photometry and spectroscopy 
for the optical counterparts of X-ray sources detected in seven 
Einstein fields and classified 28 objects as low-mass PMS stars. 
Two large-scale surveys have been performed recently. The first 
was conducted by Kunkel (1999) who observed optical counterparts 
of more than 200 {\it{ROSAT}} All-Sky Survey (RASS) X-ray sources 
in a $\sim60 \rm{deg}^2$ area in Upper-Scorpius and Upper 
Centaurus-Lupus. The other study was a spectroscopic survey for 
PMS stars in a $160 \rm{deg}^2$ area by Preibisch et al. (1998). 
A number of further searches have been performed, all focused on 
small subregions within the association (Meyer et al. 1993; 
Sciortino et al. 1998; Mart\'\i n 1998). A study of the history of star formation in Upper-Sco has been 
published by Preibisch \& Zinnecker (1999). The authors obtained 
R and I photometry of association members to about $I\sim 12.8$ 
and $R\sim 14$. 

With the goal of extending the low-mass sequence of the association, 
we have obtained photometry for the association in the R, I, and Z 
filters. We have also observed selected member candidates in the J and
H filters and spectroscopically. Our search starts at $I\sim13$ and it is
therefore complementary to Preibisch \& Zinnecker (1999). 
Section 2 describes the observations and section 3 outlines the results.

\section{Observations}

A summary of all observations is in Table 1.

We obtained photometry for eight fields of 80 by 80 arcmins in 
Upper-Sco using the 60cm Michigan Curtis-Schmidt telescope at CTIO. 
We therefore cover $\sim10\%$ of the association. Figure 1 shows the 
location of the fields. We observed the fields in the R, I, and Z 
filters. Raw frames were reduced within the IRAF\setcounter{footnote}{0}\footnote{IRAF is distributed by National Optical Observatories, which is operated by the Association of Universities for Research in Astronomy, Inc., under contract with the National Science Foundation} environment, using the CCDRED package. The images were bias-subtracted and flat-fielded. The photometry was obtained using the PSF fitting routines from IRAF. As stars are undersampled (the average FWHM of a star is $\sim1.5$ pixels) the errors in the photometry are dominated by centering errors. We obtained magnitudes of more than 180,000 
stars. 

The completeness limits for each filter are: $R\sim19$ 
magnitudes, $I\sim18.5$ magnitudes, $Z\sim18.5$ magnitudes. 
All our fields saturate at $\sim13$ mag. in all the filters.
Figures 2ab show the color-magnitude diagrams for R-I and I-Z. 
Also shown are the completeness limits and saturation limits. 
We select as preliminary candidates those 
objects that lie to the right of the Leggett (1992) main sequence 
in both color-magnitude diagrams and below the saturation limit in the I vs. I-Z color-magnitude diagram. This assumption will most certainly increase the number of candidates that are not cluster members. The best way of doing the selection would be to 
include the errors due to photometry and undetected binaries in 
the estimation of the number of objects in a band around the 5 Myr isochrones. However, given 
that we do not have {\it{a priori}} information about reddening, it seems safer to begin by assuming that every object to the right of the Leggett (1992) 
sequence will be a member of the cluster. 

With this selection method we also take account
of the saturation of the R-I color for low-mass stars. As has been 
shown by Bessell (1991), the R-I color saturates at $R-I\sim2.4$ 
(due to molecular absorption in the stellar photosphere) and becomes 
bluer for cooler objects. Therefore, a selection based only on a 
band around the 5 Myr isochrones would miss the very low-mass 
objects. Figure 3 shows the I vs. 
R-I color-magnitude diagrams for the resulting candidates. As can be seen from Figure 3, the saturation limit in the Z filter affects the selection of candidates brighter than $I\sim13.5$ which corresponds to $\sim M2$. In the lower limit, we will miss objects with $R-I>2.2$.

We obtain 138 candidates, listed in Table 2 (finding 
charts can be obtained by contacting the authors). If all candidates belonged to the cluster then we would neatly cover all the range 
of M stars. As mentioned in the introduction, Preibisch \& Zinnecker (1999) obtained R and I photometry of association members to about 
$I\sim 12.8$ and $R\sim 14$. Our search therefore begins where 
theirs ended.

We have complemented the optical observations with infrared J and 
H observations of selected candidates. To perform these observations 
we used the Cerro Tololo Infrared Imager (CIRIM) at CTIO. Raw frames were reduced within the IRAF environment, following the procedure outlined by Joyce (1992).  We were able to
observe only 9 objects, due to bad weather and instrumental problems. 
Table 3 details the results.

Using the red arm of the KAST spectrograph at the Lick 3m telescope 
 (which covers the range from 5000 to 10000 \AA), we observed selected bright candidates in low resolution (Grating 300/7500, which gives
 $\bigtriangleup\lambda\sim 11$ \AA \ of resolution). Raw images were reduced within the IRAF environment, using standard tools to perform flat-fielding, optimal extraction, wavelength calibration and response correction of the spectra. The spectra were not corrected for telluric absorption. 

Because of the low declination of the cluster, 
observations from Lick Observatory must be made through a large 
airmass. This, together with the low resolution of the
 observations and the variable fringing of the spectrograph in the
 red arm, make it very difficult to comment on gravity 
sensitive lines that lie on regions of telluric absorption, such as
 the K I resonance doublet ($\lambda\lambda7665,7699$) or the subordinate lines of the Na I doublet ($\lambda\lambda8183,8195$). However, the spectral 
resolution is good enough to identify whether or not a star has 
\hal in emission. We have used the I3 index defined by Mart\'\i n \& Kun
 (1996) and the VO index defined by Kirkpatrick, Henry \& Simons 
(1995) to find the spectral type of the Lick stars. These indexes have the advantage of being based on flux ratios that are close to each other. Therefore, they are not very sensitive to reddening. The spectral type obtained from the spectroscopic indexes can be compared with the spectral type from the colors to obtain the reddening. The results 
are shown in Table 3. The spectral types derived from the spectroscopy confirm the spectral types from the colors. 

\section{Discussion}

\subsection{Contamination by other sources}

Photometric observations as a way to select cluster candidates
 are susceptible to contamination by foreground and background 
objects. There are four kinds of objects that may appear above 
the main sequence in the color-magnitude diagram: background 
giants, background galaxies, reddened background stars, and 
foreground low-mass stars. 

From the results of Kirkpatrick et al. (1994) it is possible
 to show that the number of contaminating giants is negligible 
($<5\%$ for all spectral types). Contamination by galaxies is 
not important for the region of interest in the color magnitude
 diagram (Bouwens et al. 1998ab).

To estimate the contamination due to reddened background 
stars we use the maximum reddening towards the Upper-Sco region,
 which is $A_V\sim2$ (Schlegel, Finkbeiner \& Davis, 1998). 
Using the density of background stars (from the observed population 
of background stars under the Leggett sequence in the color-magnitude 
diagrams) we find that at most 25 candidates may be background stars:
 7 before M4, 10 between M4 and M5, 8 after M5. These will 
preferentially lie close to the Leggett sequence.

Another source of contamination is foreground field 
M-stars in the cluster line of sight. Using the luminosity function
 derived by Kirkpatrick et al. (1994) we find that there should be 
10 field stars between M4 and M5, 3 between M5 and M6, 6 between M6 
and M7, and 15 between M7 and M9. Before M4 we expect a lower limit 
of 15 field stars.

\subsection{Spectroscopic observations}

Low resolution spectroscopic observations of 22 candidates were made with the idea of determining cluster membership. As was mentioned before, the observations are not detailed enough to give information about gravity, but they can provide information about activity in the form of the \hal line. As has been shown by Prosser, Stauffer \& Kraft (1991), activity decreases with age, and therefore a strong \hal line in emission is an indicator of youth. Prosser et al. (1991) and Liebert et al. (1992) have also shown that activity increases with spectral type, starting about M1 for field dwarfs. The \hal equivalent width reaches $-12$ \AA \ at about M9.

We find \hal in 20 of the 22 objects observed spectroscopically. The values of the equivalent width are in Table 3. In Table 3 we also compare the spectral types determined from the spectra to those determined from CIRIM observations (I-J) when possible. As the Table shows, the results between the measurements are consistent, which confirms the accuracy of the photometry. Figure 4 shows the traces of five representative spectra, uncorrected for telluric absorption but corrected for reddening. 

Figure 5 shows the comparison of the measured \hal equivalent widths in Upper-Sco with those in the $\sigma$-Orionis (3-7 Myr, B\'ejar et al. 1999)
 and the $\alpha$-Persei (60 - 90 Myr, Stauffer et al. 1999, Basri \& Mart\'\i n 1999) clusters. Also shown is the envelope of \hal equivalent widths for field stars (Prosser et al. 1991; Liebert et al. 1992).  Overall, the Upper-Sco values lie above those of $\alpha$-Per and below those of $\sigma$-Ori. 9 of the Upper-Sco objects are below the envelope defined by field stars, which means that their \hal strength is consistent with them not being cluster members. However, the smoothness of the envelope is misleading. The original data for field stars shows that the maximum \hal equivalent width for each spectral type has a fair amount of scattering. Furthermore, spectroscopic observations of the young $\sigma$-Orionis cluster have shown that the three objects below the field-star envelope are likely to be cluster members. These arguments show that the presence of \hal alone is not enough to confirm or deny the status of a candidate as a cluster member. 

Figure 6 shows the color-magnitude diagram for the objects 
observed spectroscopically. All the colors and magnitudes have been corrected for reddening, if known from the spectroscopy. The errors in each axis are $\sim \pm 0.1$ magnitudes. We have included in Figure 6 the isochrones calculated by D'Antona \& Mazzitelli (1994). The translation from their Luminosity-T$_{eff}$ calculations to I and R-I involves a color-T$_{eff}$ scale and bolometric corrections. We have used the transformations given by Bessell, Castelli \& 
Plez (1998) and the colors for low-mass stars from Kirkpatrick \& 
McCarthy (1994). As Bessell et al. (1998) give bolometric 
corrections only for dwarf stars and the Kirkpatrick \& McCarthy 
(1994) colors are from field stars, the color-magnitude isochrones in 
Figure 6 suffer from considerable uncertainties. This is another 
reason to base the selection of candidates in terms of the Leggett 
(1992) main sequence.

Besides D'Antona \& Mazzitelli (1994), Burrows et al. (1997) and Baraffe et al. (1998) have published low-mass isochrones. Burrows et al. (1997) provide luminosities and effective temperatures, and so the translation to observable quantities suffer from the same problems as indicated above. The predicted masses for each color are smaller by about 50\%, compared to those of D'Antona \& Mazzitelli (1994). Baraffe et al. (1998) provide isochrones in the color-magnitude space but they are too blue, touching the observed Leggett main sequence already at 5 Myr. A comparison of the models of the three groups can be found in B\'ejar, Zapatero-Osorio \& Rebolo (1999). This comparison shows that there may still be systematic errors in the R,I isochrones.

Even taken into account the measurement errors, the objects are scattered around various isochrones. Assuming they all belong to the cluster, it is not clear from these observations what would be the correct age of the association, because even those objects with strong \hal (e.g. objects above the \hal envelope for the field) do not all fall on a single isochrone.  If one believes the 5 Myrs estimate for the age of the cluster (Preibisch \& Zinnecker 1998), the scatter has to be explained by other means. If some of the spectroscopic objects were unresolved binaries, their magnitudes would have to be increased. This would work for the latest types of objects, but not for earliest types. As mentioned above, errors in the theoretical isochrones are also a possibility. However, it does not seem likely that any realistic adjustment in the models will make all the data points lie on  the same isochrone. All these arguments point to the conclusion that the scatter is probably caused by more than one factor. Without follow-up spectroscopy (see below) it is not possible to make a stronger statement.

Of the 11 objects with a spectral type between M4 and M5 we find \hal in all except two. We expect 40\% contamination in 
this bin, and given the small-number statistics involved, our 
findings of two contaminating stars are consistent with this 
 estimate, assuming that all the stars with \hal in emission are indeed cluster members. We found \hal on the eight stars observed with
 spectral types between M5 and M6. The expected contamination is 
about 10\%. The fact that we did not find any clear 
non-cluster member is consistent with the estimate. In other words, assuming that the objects with \hal in emission are cluster members is consistent with the contamination estimates.

\subsection{The Initial Mass Function}

From the D'Antona \& Mazitelli (1994) models we find that the substellar limit for this association should be at $I\sim14.6$, $R-I\sim2.1$, $\sim$M6. This is not an accurate number, given the uncertainties in the color-$T_{eff}$ scale and the bolometric corrections mentioned before. However, if one accepts this estimate, we should have 10 brown dwarfs in our spectroscopic sample, assuming that all the \hal emitters belong to the association.

One can use the Miller-Scalo IMF (Miller \& Scalo 1979) centered in 0.1 \msun to estimate the expected number of low-mass stars and brown dwarfs in our sample. Using the 34 A stars found by Hipparcos (De Zeeuw et al. 1999) we obtain that there should be 650 stars with masses between 0.07 and 0.5 \msun, and about 600 brown dwarfs with masses between 0.005 and 0.07 \msun. We are only complete to 0.03 \msun: there should be 30 brown dwarfs between 0.03 and 0.07 \msun. Given that we are covering 10\% of the cluster, we expect 65 M stars and 60 brown dwarfs, with 3 brown dwarfs having masses between 0.03 and 0.07 \msun. In the whole sample, we find $\sim20$ objects with masses between 0.03 and 0.07 \msun (the number of expected contaminants is $\sim 10$) and  $\sim90$ objects with masses greater than 0.03 \msun (the number of expected contaminants is $\sim20$). The number of M Stars (70) is similar to that predicted by the Miller-Scalo IMF. The number of brown dwarf candidates that we are finding here is a little high (3 predicted compared to 10 after taking account of the contamination). As the recent work by Paresce \& De Marchi (1999) shows, most estimations of the IMF assume a log-normal distribution, generally centered in masses larger than 0.1 \msun. Using one of such distributions would decrease even more the number of predicted brown dwarfs in our sample. 

This discrepancy between observed and predicted number of brown dwarfs may be due to an underestimation of the number of contaminants. The study by Kirkpatrick et al. (1994) mentioned in the section about contamination measures stellar populations towards the galactic poles. Such populations may not be representative of the line of sight towards Upper-Scorpius. The study by Guillout et al. (1998) finds a population of young late-type stars (down to M2) in nearby Lupus distributed along the so-called 'Gould Belt'. It is possible that the Belt population is contaminating our sample as it crosses near Upper-Scorpius, even though it is difficult to understand why it affects only the brown dwarfs range and not the low-mass stars range. Alternatively, the discrepancy between observed and expected number of brown dwarfs may reflect a real increase in the IMF at low masses, similar to that described by Bouvier et al. (1998). It is not possible to make a stronger statement about the IMF without follow-up spectroscopy of all the candidates. 

\subsection{Lithium Burning}

As models show, the transition between lithium destruction and preservation in stellar photospheres occurs at higher masses for younger clusters. At the age of Upper-Sco (5 Myr) neither very low-mass stars nor brown dwarfs have had sufficient time to reach the core temperatures necessary to start lithium burning (D'Antona \& Mazzitelli 1997; Soderblom et al. 1998). We therefore plan to search for lithium in these candidates, as they should have it if they are cluster members. As the cluster is very young, the so-called 'Lithium Test' (Magazzu, Mart\'\i n \& Rebolo 1993), the use of the substellar lithium boundary to date the cluster, cannot be applied in Upper-Sco. However, the amount of lithium in the photosphere of early M stars could be used to date the association, as an M0 member of a 10 Myr cluster should have about 4 times less lithium than if the cluster were 5 Myr old (Soderblom et al. 1998). Also, as always, an object with spectral type later than M7 and showing lithium in the photosphere should be a brown dwarf (Basri 1998). Therefore, even though there is no lithium boundary in the cluster, lithium is still an important age and mass diagnostic.

Recently, B\'ejar et al. (1999) have suggested that the deuterium boundary could be used as another way to date the cluster. For the Upper-Sco association we expect the deuterium boundary to be located at 0.04\msun \ and therefore some of our reddest candidates could still have deuterium in their atmospheres. However, detecting of deuterium abundances posses considerable challenges from the observational point of view.

UScoCTIO 128 is a very interesting object: it shows strong \hal emission, and its position in the color magnitude diagram indicates a mass of 0.02 \msun. For comparison, the models from Burrows et al. (1997) give a mass even smaller, of about 0.015 \msun. We have two measurements of the \hal equivalent width (separated by a month and taken with the same instrument in the same configuration), and the value is the same within the errors. This indicates that the strong equivalent width is probably not a flare. In this respect UScoCTIO 128 is very similar to SOri 45, found by B\'ejar et al. (1999). More spectroscopic observations are needed before we can confirm UScoCTIO 128 as the lowest mass member yet of the Upper-Sco OB association.

\section{Conclusions}

We have conducted a photometric search for the low-mass members of the Upper-Scorpius OB association, using the R, I, and Z filters. Completeness limits are R$\sim19$, I$\sim18.5$, Z$\sim18.5$. Our search covers $\sim10\%$ of the association. The photometry comfortably crosses the substellar limit for the cluster, situated at $I\sim14.6$, $R-I\sim2.1$, $\sim$M6. This is the first survey to sample the low-mass region of Upper-Sco. We find 138 candidate members of the cluster. Contamination by non-cluster members (mainly foreground M stars and background reddened stars) has been estimated to be 59 objects.

Follow-up observations using infrared images and low resolution spectroscopy confirm the optical photometry of a reduced sample of the candidates. Of 22 objects observed spectroscopically, 20 have \hal in emission, an indicator of young age. Comparisons between the \hal equivalent widths found in other clusters and those in Upper-Sco indicate that 11 of those 20 objects may be members of the association, as those 11 object have stronger \hal than expected for low-mass field stars. However, it is possible that all the 20 objects are indeed association members. The objects with strong \hal do not all fall on a single isochrone. This may be due to the presence of unresolved binaries, to contamination from field \hal emitters or to an intrinsic age spread in the cluster.

Using the Miller-Scalo IMF centered on 0.1 \msun \ we estimate the number of objects with masses between 0.07 and 0.3 \msun \ as 3 times less than what is observed. This may be due to an underestimation on the number of contaminants affecting this mass bin, due for example to a contribution of low-mass objects by the Gould Belt. On the other hand, similar excesses in the low-mass populations have been observed in other young clusters, like the Pleides, and may point to a departure of the IMF from a simple log-normal form. A stronger statement about the IMF will have to wait until we have more cluster membership diagnostics.

The spectroscopic observations do not have high enough resolution to observe gravity sensitive lines, such as the resonant transitions of K I or the subordinate lines of Na I. Therefore, they can not be used to distinguish between young low-mass objects and main-sequence objects.

We find a very interesting object, UScoCTIO 128 ($\sim$M7) with very strong \hal (equivalent width of $\sim-130$ \AA) and an estimated mass of 0.02 \msun. If this object is a member of the cluster, it would be one of the lowest-mass brown dwarfs known to date.

Full confirmation of the membership of these objects will have to wait for higher resolution spectroscopy that can observe Li I, K I, and Na I. As has been suggested by other groups, these very young objects provide a unique opportunity to study depletion of very light elements such as deuterium. Perhaps in the future studies of deuterium depletion will take the place of lithium depletion as a precise method of determining the ages of very young clusters.

\acknowledgments

We would like to thank Victor B\'ejar for invaluable help in collecting the CTIO observations and for many stimulating discussions concerning brown dwarfs. Thanks are also due to Debi Howell-Ardila, who edited the manuscript for language. We acknowledge the support of the National Science Foundation through grant number AST96-18439. EM acknowledges support from the NASA Origins program.


\clearpage

Figure 1: Location of the observed fields in Upper-Scorpius. Each field is 80' by 80'. The black dots indicate Hipparcos members (De Zeeuw et al. 1999).

Figure 2: (a) I vs. R-I color-magnitude diagram for all the objects observed. The results of the photometry are represented by small dots. The lower dashed line is the completeness limit. The solid line is the Leggett (1992) main sequence. (b) Same diagram for I vs. I-Z. The upper dashed line is the saturation limit. 

Figure 3: Color-magnitude diagram for the selected candidates. The solid line is the Leggett (1992) main-sequence. The dashed line is the completeness limit. The arrow corresponds to $A_V=1$.

Figure 4: Traces of five representative spectra with \hal. The spectra are corrected for reddening but not for telluric absorption. UScoCTIO 128 has a \hal equivalent width of -130.5.

Figure 5: Absolute values of \hal equivalent widths for various clusters. All the \hal lines are in emission. The symbols are: ($\bullet$) Upper-Sco, ($\ast$) $\alpha$-Persei, ($\diamond$) $\sigma$-Orionis. For clarity, only those objects with \hal EqW. less than $50$ \AA \ are plotted. The dashed line shows the  envelope of the \hal equivalent widths for field stars (Prosser et al. 1991; Liebert et al. 1992). The cross indicates the size of the error bars. The \hal strength in Upper-Sco is intermediate between that of $\sigma$-Orionis and $\alpha$-Persei.

Figure 6: Color-magnitude diagram of objects observed spectroscopically. All objects have been de-reddened. The dots ($\bullet$) indicate objects for which the \hal equivalent width is above the field envelope (see text). The asterisks ($\ast$) indicate objects for which the \hal equivalent width is below the field. The upper axis shows the spectral types calculated using the color-spectral type calibration from Kirkpatrick \& McCarthy (1994). Also shown are the evolutionary models by D'Antona \& Mazzitelli (1994), for masses from 0.20 to 0.02\msun, and isochrones from 1 Myr to 10 Myr Of those objects observed spectroscopically only two (UScoCTIO 28 and UScoCTIO 162) do not have \hal.


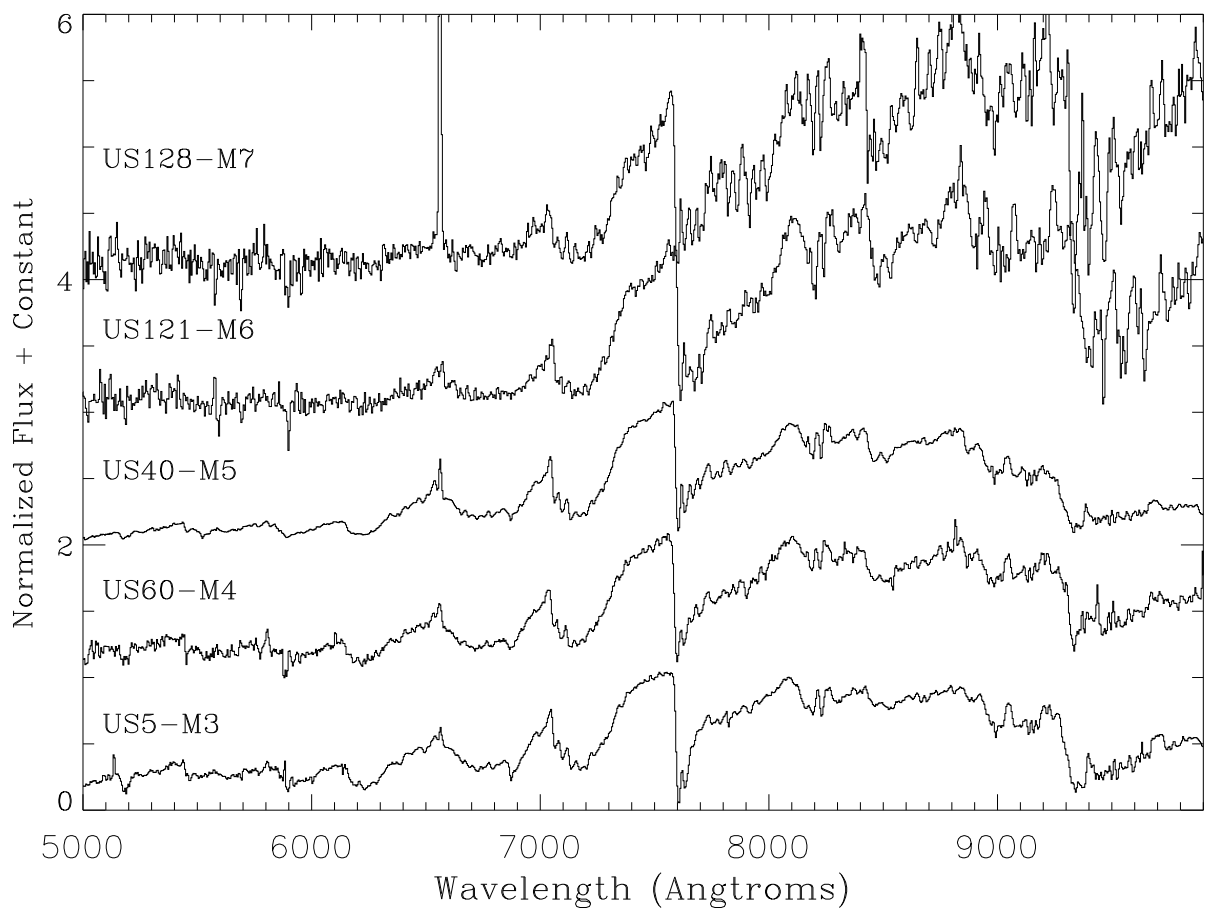
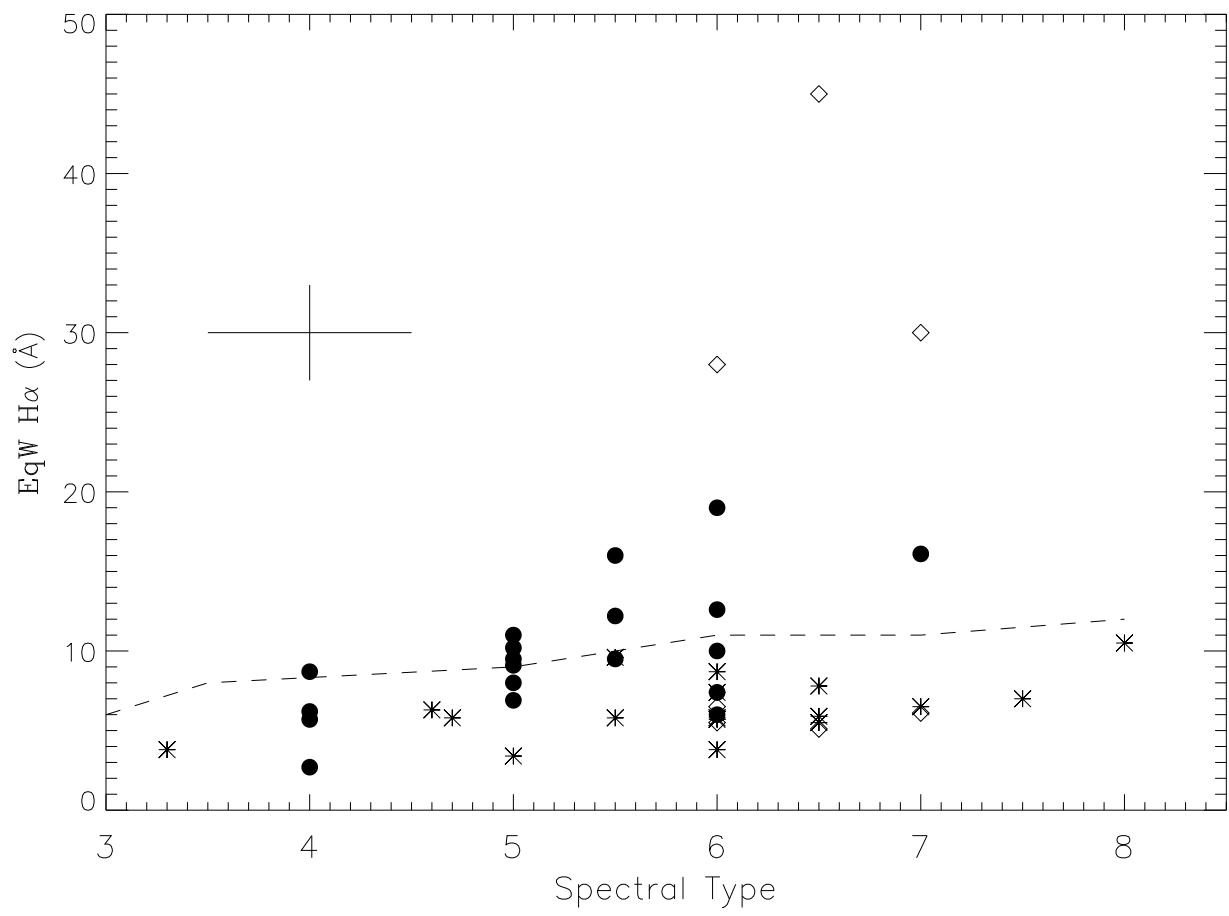
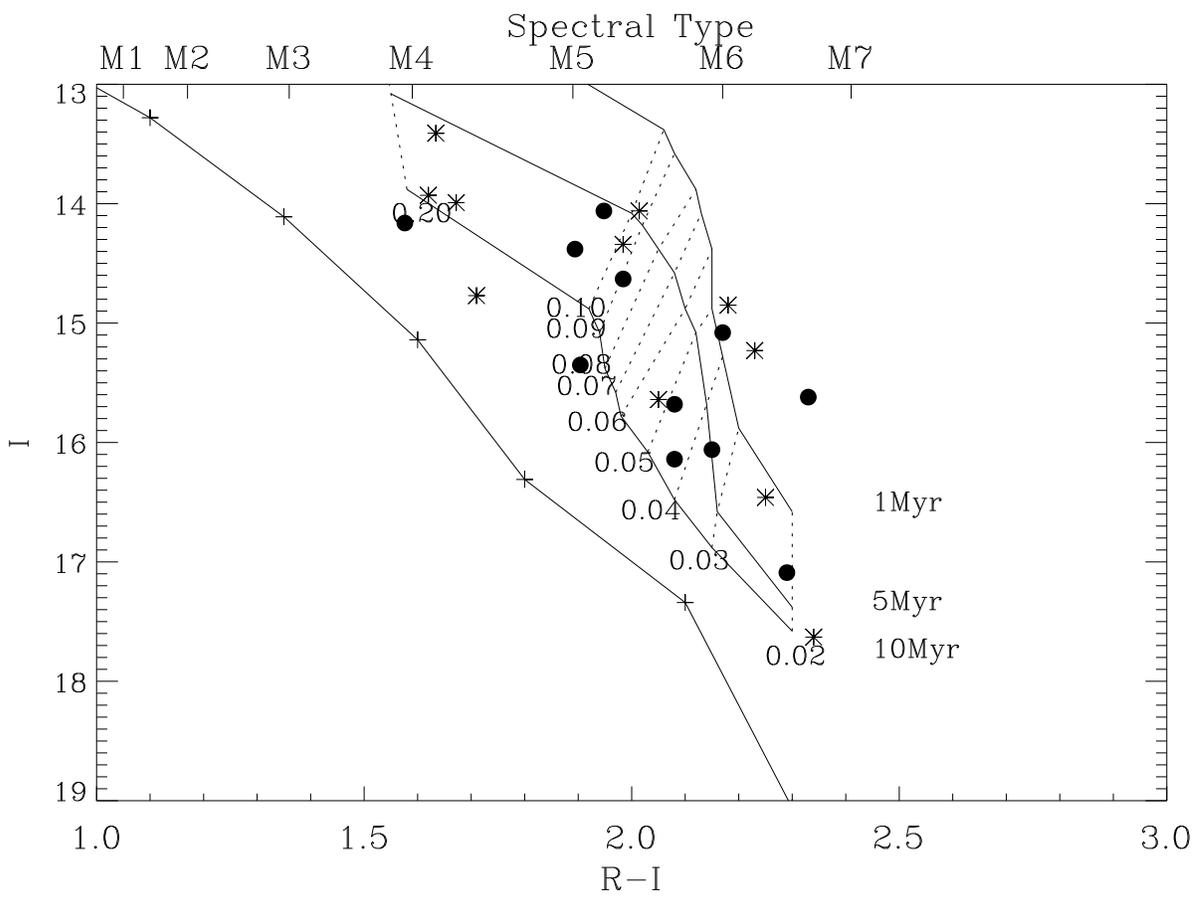
\end{document}